\documentclass[preprint,aps,prb]{revtex4}
\usepackage[dvips]{graphicx}
\usepackage{amssymb}

\def\psiav{\langle\psi\rangle}
\def\Eav{\langle E\rangle}
\def\Peq{P_\mathrm{eq}}
\def\Geq{G_\mathrm{eq}}

\def\hpsi{\hat{\psi}}

\def\ett{\epsilon_{TT}}
\def\eth{\epsilon_{TH}}
\def\ehh{\epsilon_{HH}}
\def\Emin{E_\mathrm{min}}
\def\Emax{E_\mathrm{max}}
\def\tinner{t_\mathrm{inner}}

\begin{document}

\title{Coarse-Grained Kinetic Computations for Rare Events:
	Application to Micelle Formation}
\author{Dmitry I. Kopelevich}
\altaffiliation[Current address: ]{Department of Chemical Engineering,
	     University of Florida,
             Gainesville, FL 32611}
\email{dkopelevich@che.ufl.edu}
\author{Athanassios Z. Panagiotopoulos}
\email{azp@princeton.edu}
\author{Ioannis G. Kevrekidis}
\thanks{Corresponding author}
\email{yannis@arnold.princeton.edu}
\affiliation{Department of Chemical Engineering\\
Princeton University\\
Princeton, NJ 08544}
\begin{abstract}
We discuss a coarse-grained approach to the computation of rare events
in the context of grand canonical Monte Carlo (GCMC) simulations of 
self-assembly of surfactant molecules into micelles.
The basic assumption is that the {\it computational} system dynamics 
can be decomposed into two parts -- fast (noise) and slow 
(reaction coordinates) dynamics, so that the system can be described by 
an effective, coarse grained Fokker-Planck (FP) equation. 
While such an assumption may be valid in many circumstances, an 
explicit form of FP equation is not always available. 
In our computations we bypass the analytic derivation of such an
effective FP equation.
The effective free energy gradient and the state-dependent magnitude of the random noise, 
which are necessary to formulate the effective
Fokker-Planck equation, are obtained from ensembles of short bursts of 
microscopic simulations {\it with 
judiciously chosen initial conditions}.
The reaction coordinate in our micelle formation problem is taken to 
be the size of a cluster of surfactant molecules.
%
%
We test the validity of the effective FP description in this system and 
reconstruct a coarse-grained free energy surface in good agreement with full-scale 
GCMC simulations. 
We also show that, for very small clusters, the cluster size seizes to
be a good reaction coordinate for a one-dimensional effective description.
We discuss possible ways to improve the current 
model and to take higher-dimensional coarse-grained dynamics into account.
\end{abstract}
\maketitle

\section{Introduction}
The development of efficient computational methods for the study of rare events is
a subject of intense current interest and research across many disciplines 
\cite{Kapral89,Voter00,Chandler98,Chandler99,Parrinello02,WeiCai03}.
Direct microscopic (e.g., molecular dynamics or kinetic Monte 
Carlo) simulations of rare events can be extremely time-consuming 
since most of the computational
time is spent sampling the free energy surface close to local free energy minima and 
the transition states are sampled only during an exponentially small fraction of the 
simulation time. 

Many systems can be successfully described through a low-dimensional 
{\it effective free-energy surface} in terms of dynamically meaningful
observable quantities (often referred to as ``reaction coordinates",
see for example the discussion in Ref.~\onlinecite{HK03}).
In such cases it is reasonable to expect that 
the evolution of the probability density of the variables that parameterize
this surface may be described in terms of an effective Fokker-Planck (FP)
equation.
The deterministic part of the FP equation will then contain the gradient of the 
effective free energy surface with respect to the few ``coarse" variables (reaction coordinates,
``observables") chosen to parameterize it, as well as the local diffusivity
of the expected motion. 

In many cases of practical interest, this effective FP equation is not available
in closed form. 
Recently, Hummer and Kevrekidis~\cite{HK03} have proposed a so called
kinetic approach that bypasses the analytical derivation of 
such an equation, and uses the concept of its existence to guide the detailed
(molecular dynamics, Monte Carlo) simulations. 
In this approach, the components
of the effective FP equation are estimated through multiple, 
relatively short microscopic simulations with judiciously
chosen initial conditions.

In this paper, we apply this kinetic approach to Monte Carlo (MC) simulations of 
self-assembly of surfactant molecules into micelles.
We consider a lattice model \cite{LarsonMC85,LarsonMC96} with only 
short-range hydrophobic interactions between the molecules. 
The ``dynamics" of these simulations are artificial;
the kinetic approach allows us to explore the free energy
surface using this artificial dynamic evolution. 
%
%
Indeed, in section~\ref{S:mix}
we show that the free energy surface predicted by the kinetic
approach does not depend on a particular choice of the  MC ``dynamics". 
Moreover, we expect that the methods and conclusions of this work
can also be applied to real-time, (molecular-dynamic) simulations
of self-assembly.

The assumption of an effective-FP (and associated Langevin) dynamics of micelle formation 
is a departure from the usual assumption of the first order activated 
process of addition (removal) of single amphiphile molecules to (from) a micelle 
\cite{Aniansson74,Neu02}. 
However, we show that the effective Langevin equation model performs well 
for the considered system; this suggests a link
between the effective Langevin description and the 
master equation of the activated process model. 
This link needs to be investigated in the future.

In the companion paper \cite{KPK04a} (Paper I), we have 
considered ``dynamics" of MC simulations of micelle formation
and have discussed the application of the multiscale coarse projective integration
and coarse Newton methods to these systems. 
We have observed that, as in the real physical dynamics, the rate-limiting
step in the MC ``dynamics" is the birth and death of micelles (as opposed to, e.g.,
altering of micelle shape and size as the external parameters are changed).
In Paper I, we have used the first two moments of the micelle number
density as our coarse variables (reaction coordinates,``observables").
In addition, we have performed the coarse computations of the
system using a kinetic Monte Carlo (kMC) model for micelle 
birth and death with the rates
obtained from the full-scale equilibrium simulation.

In the current paper, we compute the micelle formation rates
directly from short-time MC simulations
using the kinetic approach.
The coarse variables here are the variables that characterize
micelle size and shape.
%
We assume that the coarse variables that can be used to
parameterize a free energy surface description can be selected among 
physical attributes (such as size, energy, radius of gyration) 
of a cluster of surfactant molecules.
Based on computational evidence supporting the 
existence of effectively one-dimensional long-term dynamics, 
we use a single coarse degree 
of freedom (a single ``reaction coordinate", the cluster size) 
to parameterize the effective free energy surface and show that 
the remaining coarse degrees of freedom relax quickly to functions of (become slaved to)
a single ``master mode". 
We then demonstrate the validity of assumptions of the effective FP
dynamics for the cluster size and estimate 
the effective free energy surface as well as the rates of micelle breakup, 
in good agreement with full equilibrium MC simulation.

We observe that the assumption of one-dimensional coarse dynamics breaks down for 
small cluster sizes and that, in order to successfully reconstruct the 
free energy surface, one needs to consider coarse-grained dynamics in at least a
two-dimensional configuration space, where the second dimension can be chosen
to be, e.g., the cluster energy.

The paper is organized as follows: In Section \ref{S:theo},
we state some basic results of the theory of stochastic processes, 
which form theoretical basis of the coarse kinetic 
approach. 
In Section \ref{S:equilibrium}, we briefly review the lattice 
model and the Monte Carlo method used in our simulations. 
We also present results of a long-time equilibrium simulation, 
which will be compared with the kinetic approach results in the 
subsequent sections. 
Section \ref{S:nonequil} contains a detailed description of our
implementation of the kinetic approach specific
to simulations of micelle formation.
Results of the kinetic approach calculations are reported
in Section \ref{S:results}. 
In this section, we also check 
assumptions underlying the effective FP equation model for the 
micelle formation dynamics and validate these assumptions computationally.
In Section \ref{S:multi_dim}, we explore the micelle formation
dynamics in the phase space parameterized by two coarse
variables. 
We observe that, in most cases, the system quickly approaches an effective
one-dimensional manifold -so that the one-dimensional FP model
for the micelle formation dynamics is appropriate. 
We further observe
that such a separation of timescales is much weaker for dynamics
of small clusters.
Finally, in Section \ref{S:Discussion}, we summarize our findings
and briefly discuss some other coarse-grained, ``equation-free"
 methods (coarse Newton and coarse reverse
integration) and their application to the micelle formation
problem.

\section{Theoretical background.}
\label{S:theo}

In this section we review some standard results from the
theory of stochastic processes and discuss their 
role in the kinetic approach.
It is assumed here that the system dynamics can be described by
a single coarse variable (reaction coordinate) $\psi(t)$.
This assumption implies that all other variables 
quickly approach some sort of slow, attracting, one-dimensional manifold;
that is, the statistics of the simulation quickly become functions
of one observable; the slow manifold is the graph of this function.
In our case
of micelle formation, $\psi$ is chosen to be the size of a micelle,
as measured by the number of amphiphile molecules contained in the 
micelle and it is assumed that all other physical attributes of a micelle
(such as radii of gyration, density profile, energy, etc.) are 
quickly slaved (in an averaged sense) to its size. It will be shown in Section~\ref{S:multi_dim} that this is a reasonable assumption.

Consider a general one-dimensional stochastic process $\psi(t)$.
The evolution of the probability density $P(\psi,t)$ of $\psi$
obeys the following integral equation \cite{Risken}
\begin{equation}
   P(\psi,t+\tau) = \int \rho(\psi,t+\tau\vert \psi',t) P(\psi',t) d\psi',
\end{equation}
where $\rho(\psi,t+\tau\vert \psi',t)$ is the transition probability
from point $\psi'$ at time $t$ to point $\psi$ at time $(t+\tau)$.
The differential form of this equation, known as the 
Kramers-Moyal expansion, is as follows:
\begin{equation}
\label{eq:Kramers_Moyal}
   \frac{\partial P(\psi,t)}{\partial t} = 
   \sum_{n=1}^\infty \left( -\frac{\partial}{\partial \psi}
   \right )^n
   D^{(n)}(\psi,t) P(\psi,t),
\end{equation}
where 
\begin{equation}
\label{eq:Dn_def}
   D^{(n)}(\psi,t) = \frac{1}{n!} \lim_{\tau\to 0}
                  \frac{1}{\tau} \langle (\xi(t+\tau)-\xi(t))^n \rangle 
                  \vert_{\xi(t) = \psi}
\end{equation}
are the differential moments of the transition probability $\rho$.
The angular brackets here denote ensemble averaging and $\xi$ denotes
a realization of the stochastic process with a $\delta$-function distribution
at the starting point $t$, $\xi(t) = \psi$.

This is a very general result and it applies to any
one-dimensional stochastic process. 
If the process is Markovian and Gaussian, then only the first two
terms in Eq. (\ref{eq:Kramers_Moyal}) are non-zero. Moreover, if the
stochastic process is invariant with respect to the shift in time
(which is true for the processes without external time-dependent forcing),
then the expansion coefficients $D^{(n)}$ are independent of time.
Hence, under these assumptions the stochastic process can 
be described by the Fokker-Planck equation \cite{Risken}
\begin{equation}
\label{eq:FP}
   \frac{\partial P(\psi,t)}{\partial t} = 
   \left [ -\frac{\partial}{\partial \psi} v(\psi) 
   +\frac{\partial^2}{\partial \psi^2} D(\psi) \right ] P(\psi,t).
\end{equation}
Here, $v(\psi) \equiv D^{(1)}(\psi)$ is the drift coefficient and 
$D(\psi) \equiv D^{(2)} (\psi)$ is the diffusion coefficient which are 
directly related to the \emph{short-scale} evolution of the first two
moments of $\psi$ via Eq. (\ref{eq:Dn_def}).

This, in turn, implies that the Fokker-Planck equation components
(the drift and the diffusion coefficient)
can be fully reconstructed from
short-scale simulations.
For our coarse-grained dynamics, we initialize the system consistently
with some value of the coarse variable $\psi_0$ 
(we call this procedure of constructing microscopic initial
conditions consistent with the prescribed coarse variables
as ``lifting" \cite{manifesto}).
Then we perform
a short-scale simulation and estimate the derivatives of the average
and the standard deviation of the coarse variable,
\begin{equation}
\label{eq:vd_def}
   v(\psi_0,t) = \frac{\partial\langle \psi(t,\psi_0) \rangle}
                      {\partial t}, \quad
   D(\psi_0,t) = \frac{1}{2} \frac{\partial \sigma^2(t,\psi_0)}
                                  {\partial t}.
\end{equation}
Here, $\psi(t;\psi_0)$ is a trajectory of the system that
starts from $\psi = \psi_0$ at time $t=0$, angular brackets 
denote averaging over different realizations of this trajectory,
and $\sigma^2(t;\psi_0)$ is the variance of $\psi(t;\psi_0)$.
Hence, we can reconstruct a global PDE from short-scale,
appropriately initialized local simulations.
In practice (in this paper) 
the derivatives contained in expressions (\ref{eq:vd_def})
are computed by fitting a straight line to $\psiav(t)$ and $\sigma^2(t)$.
This procedure is discussed in more detail in Section~\ref{S:nonequil};
clearly, better fitting techniques (e.g. maximum likelihood estimation)
can be used.
Once the Fokker-Planck equation is reconstructed, one can calculate 
several global characteristics of the system, such as the effective free energy $G(\psi)$ 
and the rates of transitions between different metastable states of the system. 
This effective free energy can be obtained from the 
equilibrium probability distribution $\Peq(\psi)$ which, in turn, is
a solution of the steady-state Fokker-Planck equation
\begin{equation}
\label{eq:SFP}
   \left [ -v(\psi) + \frac{\partial}{\partial \psi} D(\psi) \right ] 
   \Peq(\psi,t) = 0.
\end{equation}
By substituting the ansatz $\Peq(\psi) \propto \exp(-G(\psi)/k_B T)$
into the equation (\ref{eq:SFP}), we obtain
\begin{equation}
\label{eq:G}
  G(\psi) = - k_B T\int \frac{v(\psi')}{D(\psi')} d\psi' + k_B T \ln D(\psi)
          + const.
\end{equation}
Here, $k_B$ is the Boltzmann factor and $T$ is the temperature of the
system. 
Note that, since the free energy is defined up to an additive
constant, one can multiply $D(\psi)$ in the logarithmic term by an arbitrary
constant in order to preserve consistent units.
The second term of Eq. (\ref{eq:G}) is significant if the noise
is multiplicative.
Since this is a subtle point that can be overlooked if one uses other
(equivalent) descriptions of the stochastic process, we here discuss
it in more detail.

Let us first discuss the connection between the Fokker-Planck equation and the
corresponding Langevin equation descriptions.
This point would become important if we tried to
fit simulation data to a coarse-grained {\it Langevin} description -rather than a
coarse-grained FP description.

A Langevin equation that corresponds to the Fokker-Planck equation (\ref{eq:FP})
is as follows:
\begin{equation}
\label{eq:LE}
   \dot{\psi} = \frac{1}{\gamma(\psi)} (f_0(\psi) + \Gamma(t) )
\end{equation}
Here, $\gamma(\psi)$ is the friction coefficient,
$f_0(\psi)$ is the deterministic force, and $\Gamma(t)$ is the stochastic force.
The latter is a Gaussian stochastic process with zero mean and with
variance related to the damping coefficient $\gamma$ by the 
fluctuation-dissipation theorem:
\begin{equation}
\label{eq:FDT}
   \langle \Gamma(t) \Gamma(t+\tau)\rangle = 2 \gamma k_BT\delta(\tau).
\end{equation}

The relationship between $f_0(\psi)$ and $\gamma(\psi)$ of the Langevin equation
and the drift and diffusion coefficients $v(\psi)$ and $D(\psi)$ of 
the Fokker-Planck equation
depends on the interpretation of the white noise in the Langevin equation
(\ref{eq:LE}) as discussed in standard references (see e.g. Refs.~\onlinecite{Risken,Gardiner}). 
If one uses It\^o interpretation, then 
\begin{eqnarray}
\label{eq:Ito}
   v(\psi) & = & f_0(\psi)/\gamma(\psi), \\
   D(\psi) & = & k_B T/\gamma(\psi),
\end{eqnarray}
and, if one use the Stratonovich interpretation, then
\begin{eqnarray}
\label{eq:Stratonovich}
   v(\psi) & = & f_0(\psi)/\gamma(\psi) - 
              \frac{\gamma'(\psi)}{2\gamma^2(\psi)}k_B T, \\
   D(\psi) & = & k_B T/\gamma(\psi).
\end{eqnarray}
Both of these interpretations are identical
in the case of additive noise 
(i.e. when $\gamma$ is a constant independ on $\psi$).
In the case of multiplicative noise
(i.e. when $\gamma$ is depends on $\psi$),
the situation becomes
somewhat more complicated and, in particular, the force $f_0(\psi)$
is not just a gradient of the free energy $G(\psi)$ for both
It\^o and Stratonovich interpretation. This can be confirmed by
direct substitution and is discussed in detail in 
Ref.~\onlinecite{Arnold_LE00}.
In particular, in the case of It\^o interpretation,
\begin{equation}
   f_0(\psi) = -G_0^\prime(\psi),
\end{equation}
where
\begin{equation}
\label{eq:G0}
   G_0(\psi) = - k_B T\int \frac{v(\psi)}{D(\psi)} d\psi + const.
\end{equation}
It is clear that the expressions (\ref{eq:G}) and (\ref{eq:G0}) are 
identical up to an additive  constant \emph{only}  
if the diffusion coefficient $D$ (and hence the damping
coefficient $\gamma$) is constant.
Therefore, we compute the effective free energy using Eq. (\ref{eq:G}).

Despite the fact that Eq. (\ref{eq:G0}) is
an incorrect expression {\it for the free energy}, 
the quantity $G_0(\psi)$ finds its use in calculations
of transition rates.
In fact, the mean residence time in a free energy well can 
be written as \cite{Gardiner}
\begin{equation}
\label{eq:mean_res}
  \tau = \int_{\psi_0}^\psi dy\ e^{G_0(y)/k_BT} 
            \int_y^\infty dz\ e^{-G_0(z)/k_BT}{D(z)},
\end{equation}
where $\psi$ is a point inside the well, $\psi_0$ is the boundary of 
the well. 
%
This expression is used in section~\ref{S:nonequil}
to compute the micelle (computational) disintegration rate. 
If the free energy barrier is sufficiently high, then 
transitions such as micelle formation and disintegration,
can be described by 
first order kinetics and the transition rate $k$ is the inverse
of the mean residence time $\tau$.

Earlier work on such a kinetic approach \cite{HK03} has
used the Langevin equation description of the stochastic process.
The information about the system dynamics
was obtained from the time derivatives of $\psiav(t)$ and $\sigma^2(t)$
which, in turn, were obtained by fitting $\psiav(t)$ and $\sigma^2(t)$
to a straight line,
just like for the FP equation description, see Eqs 
(\ref{eq:vd_def}).
Therefore, the \emph{fitting} procedure for 
the Langevin equation model is the same as that for the FP equation
model. However, if the diffusion coefficient is not constant,
the \emph{interpretation} of the fitting results 
for the Langevin equation can lead to ambiguities since, in this case,
one would have to specify an interpretation of the white noise
(It\^o or Stratonovich).
%
%
We will bypass here the details of the estimation (fitting) of the
data to a Langevin description that arise from the interpretation dilemma,
and use the
Fokker-Planck description of the stochastic process which directly
relates the fitted drift and diffusion coefficients to the statistical
properties of the process.

Another popular description of a stochastic process is the Smoluchowski
equation
\begin{equation}
\label{eq:Smoluchowski}
   \frac{\partial P(\psi,t)}{\partial t} = \frac{\partial}{\partial \psi}
   D(\psi) \left [ -\frac{f(\psi)}{k_BT}  + 
   \frac{\partial}{\partial \psi} \right ] P(\psi,t),
\end{equation}
which was originally derived from a 
Fokker-Planck equation for an inertial Brownian
particle in the limit of negligible inertia
\cite{Risken}. 
The advantage of the Smoluchowski equation is that $f(\psi)$ is
the ``true" effective force, i.e. $f(\psi) = -G'(\psi)$. 
However, 
$f(\psi)$ of the Smoluchowski equation, in general, is not
proportional to the drift coefficient $v(\psi)$ discussed
earlier.
In fact, some straightforward algebra shows that
\begin{equation}
   -G'(\psi) = f(\psi) = \frac{v(\psi) - D'(\psi)}{D(\psi)} k_B T,
\end{equation}
which is consistent with Eq. (\ref{eq:G}).
Hence, the correction due to the position-dependent diffusion coefficient
(the second term in Eq. (\ref{eq:G})) is present also in the Smoluchowski equation.
It will be shown in Section~\ref{S:nonequil} that
this correction
is significant in the case of the micelle formation, where
the diffusion coefficient is significantly inhomogeneous.

\section{Model and equilibrium simulation details.}
\label{S:equilibrium}

We study the micellization process using
the lattice model for surfactant systems 
originally proposed by Larson \cite{LarsonMC85,LarsonMC96}. 
Panagiotopoulos and coworkers \cite{AZPmicelles99,AZPmicelles02a}
have performed
extensive grand canonical Monte Carlo (GCMC) simulations of
this model in order to study micellization and phase transitions.
Despite its simplicity, this model yields predictions that are in
good qualitative agreement with experimental data.

In this model, an amphiphile molecule is represented as a
chain of beads and
a solvent molecule is represented as a single bead.
The beads occupy sites on a cubic lattice and the
connected beads of an amphiphile molecule are 
restricted to be in  
nearest-neighbor sites with bonds along the vectors 
$(0,0,1)$, $(0,1,1)$, $(1,1,1)$ and their reflections along
the principal axis, resulting in a coordination number of 26.
There are 
two types of beads: hydrophobic tail (T) and hydrophilic
head (H) and the solvent is modeled by head beads.

The hydrophobic interaction is modeled by attractive interaction
between tail beads.
Each bead interacts only with the 26 nearest neighbors and the
total energy of the system is the sum of pairwise interactions between 
beads.
The tail-tail interaction energy $\ett$ is -2 and the tail-head
and head-head interaction
energies $\eth$ and $\ehh$ are zero, following 
Ref.~\onlinecite{AZPmicelles02a}.
It is furthermore assumed that all sites that are not occupied
by the amphiphile beads are occupied by the solvent. 
This
assumption implies that 
there is no need to explicitly include solvent into the MC moves.

In most calculations presented in this paper, the following
mix of MC moves is used:
50\% amphiphile transfers (i.e. addition or removal), 
49.5\% amphiphile partial regrowth moves, and 0.5\% cluster moves.
In Section \ref{S:mix}, we perform simulations with several
different mixes of MC moves in order to investigate the effects 
of different ``dynamics" on the kinetic approach results.
Since in this paper we apply a dynamic approach
to equilibrium MC simulations, in order to simplify the notation, 
we refer to the number of MC move as the ``time".
Let us emphasize once again that it is the {\it MC computational}
dynamics that we attemp to -in some sense- accelerate, 
and not {\it physical} dynamics;
when the base simulation is an MD one (as in Ref. \onlinecite{HK03})
then our approach would attempt to accelerate physical dynamics.

The simulations are performed for
an amphiphile chain H$_4$T$_4$ which consists of 4 head beads 
and 4 tail beads.
The simulations are performed at temperature $k_BT = 7.0$
and chemical potential $\mu = -47.40$ in a 
cubic box with a side length of 40 sites, 
assuming periodic boundary conditions. 
This
box size is sufficient to prevent spurious effects of
periodicity, since the typical diameter of a micelle is significantly
smaller than half the size of the box side.

We perform a reference long-time simulation of the system in
order to compute the free energy 
and the rates of creation and destruction of micelles.
In this simulation, we consider 500 realizations of the system
and compute the above quantities using data saved from 600 million
MC steps after equilibration.
The free energy curve $G$ is parameterized by the cluster size
$\psi$ and is computed from the histogram of the cluster sizes.
A cluster is defined as an aggregate of amphiphile molecules
such that each molecule in a cluster has at least one tail bead which occupies
a neighboring site with a tail bead from another amphiphile of the cluster.
In other words, each cluster molecule interacts through hydrophobic attraction
with at least one other cluster molecule. 
The cluster size $\psi$ is defined
as a number of amphiphiles in this cluster.
The free energy $\Geq(\psi)$ obtained from these equilibrium
calculations is shown by the solid line in 
Fig.~\ref{F:G}a. $\Geq(\psi)$ has two minima: one at $\psi = 1$ 
which corresponds to free amphiphiles and another one at $\psi = 69$, 
which corresponds to micelles.
The free energy barrier separating these two states is located at
$\psi_b \approx 21$.

In the calculations of equilibrium micelle formation/disintegration rates,
a transition between 
a small cluster and a micelle is said to occur when the cluster
size crosses the free energy barrier $\psi_b$.
For the purposes of the equilibrium calculation, the precise 
definition of the border between
micelles and smaller clusters is unimportant since the transition
happens on the much faster timescale than the average lifetime of a
micelle.
The rate of micelle formation/disintegration or, in order words,
of transition from a system containing $i$ micelles to a 
system containing $i\pm 1$ micelles in a simulation box, is

\begin{equation}
\label{eq:kdef}
   k_{i\to i\pm1} = 1/\tau_{i->i\pm 1},
\end{equation}
where $\tau_{i->i\pm 1}$ is the average time between the transitions.
Eq. (\ref{eq:kdef}) follows from the first order kinetics approximation, 
which is justified when the time between
micelle birth/death has an exponential distribution. 
This assumption holds if the free energy barrier is sufficiently high
(which is true in the current case) 
and, moreover, we have checked this assumption by direct calculation of the
transition time distribution.
We expect that the first order kinetics assumption will break down 
in denser systems,
where the micelle coalescence becomes a dominant mechanism 
for altering size and number density of micelles.
However, as discussed in Paper I, the system under consideration 
(H$_4$T$_4$ at $k_BT = 7.0$ and $\mu = -47.40$) has low micelle density 
with an average of about 1 micelle per $40\times 40\times 40$ 
simulation box.

In the system considered here, there are no long-range energetic interactions
between the micelles. 
However, we observe that, due to entropic
interactions, the micelle
birth and death rates vary depending on the number of micelles 
already present in the simulation box. 
In this work, we focus on transitions $0\to 1$ and $1 \to 0$, i.e.
birth and death of micelles in a simulation box that is otherwise filled
only with small clusters. 
An extension to a general case of transitions
$i \to i \pm 1$ is straightforward.

\section{Details of ``kinetic" simulations.}
\label{S:nonequil}


In this section, we describe the details of the implementation of our
kinetic approach for the computational micelle formation. 
As discussed in
section \ref{S:theo}, in order to compute the drift and diffusion
coefficients, we perform short-time simulations initialized at a
prescribed value $\psi_0$ of the coarse variable $\psi$.
In the case of micelle formation, $\psi$ is the number of amphiphile
molecules contained in a micelle (or a nucleus of a micelle). 
Hence, in order to 
initialize the simulations, we place a cluster of a prescribed size
$\psi_0$ into the simulation box.

In order to facilitate this process, we maintain a database of 
cluster structures. 
In the simulations reported here, the database
is obtained from an equilibrium simulation by saving cluster structures 
every 100,000 MC steps. 
As will be shown in section~\ref{S:multi_dim},
this frequency of the database updates assures that the saved 
structures are sufficiently different from each other.
The database thus obtained contains equilibrium 
structures of clusters for some temperature $T$ and 
chemical potential $\mu$. 
In the current
paper, we consider the kinetic approach precisely for these values of $T$
and $\mu$. 
However, it is very straightforward to generate a new cluster
database from an existing one: it is only necessary to equilibrate the 
available cluster structures at new $T$ and $\mu$; we will
estimate the (relatively short) time of this equilibration
below.

Thus, the initial conditions for each simulation of the kinetic approach
consist of a single 
cluster of size $\psi_0$ picked at random from the database and placed into
an empty simulation box.
The values of $\psi_0$ range from 1 to 90 and, for each $\psi_0$, 
3000 to 10,000 MC realizations are computed.
In addition to the micelle (or a nucleus) which is explicitly placed into the
system, the system always contains some
``soup" of single amphiphiles, dimers, and other small
clusters. 
Since we do not put these small clusters into the system
explicitly, 
we let it equilibrate before computing 
statistics of the nucleus evolution.
Equilibration here means reaching a quasi-steady-state distribution of 
small clusters. 
In order to obtain the small clusters equilibration time, 
we compute the evolution of average small cluster size and 
the first two moments of the distribution of the number of molecules 
contained in small clusters and conclude that these 
quantities reach their steady-state values within 
just 20,000 MC steps. 

We hence use the equilibration time of 20,000 MC steps or more
in our simulations.
In the non-equilibrium results reported below, time = 0 corresponds to the time
at which the small clusters have equilibrated. 
%
Since the nucleus size can
change a little during the equilibration time (due to addition/removal
of amphiphiles to/from the nucleus), $\psi_0$ 
refers to the size of the cluster after the small cluster equilibration is 
complete. 
We will also discuss below the option of small cluster equilibration
{\it constrained} on the cluster size (in the spirit of umbrella sampling).

The nucleus size $\psi(t)$ is measured with some prescribed frequency 
$\Delta t$ and the center of mass of the nucleus is tracked
in order to prevent possible confusion between a small ``dying" nucleus and
an emerging small cluster.
We have performed two series of MC simulations:
      \begin{enumerate}
      \item Long simulations: Length of production
            run = $15 \times 10^4$ steps;
            frequency of output $\Delta t =  1000$ steps;
	    equilibration time before production run 
	    = $5 \times 10^4$ steps.
            
      \item Short simulations: Length of production
            run = $2 \times 10^4$ steps;
            frequency of output $\Delta t =  100$ steps;
	    equilibration time before production run 
	    = $2 \times 10^4$ steps.
           
      \end{enumerate}

The long simulations have been performed in order to study slower dynamics
of (almost) formed micelles inside
the free energy well; the short simulations have been performed 
in order to study faster dynamics near the free energy barrier as 
well as to explore the dynamics of the additional coarse variables (see sections \ref{S:results}
and \ref{S:multi_dim}). 
We observe that 
results of both simulations agree for the fast dynamics near the barrier
but the shorter simulations fail to provide sufficient information to 
correctly reconstruct the free energy surface corresponding to 
slower dynamics near the free energy minimum. 
We hence report the free energy and the diffusion coefficient 
obtained from the longer simulations.

\begin{figure}
\includegraphics[width=5in]{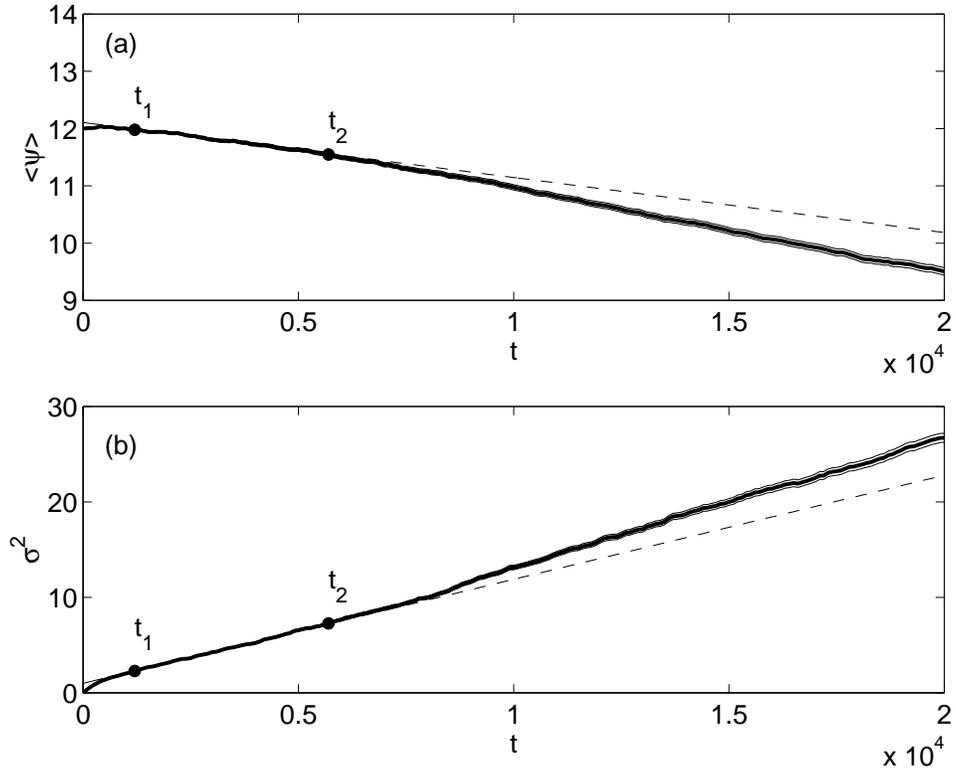}
\caption{Evolution of (a) $\langle\psi(t,\psi_0)\rangle$ and (b)
         $\sigma^2(t,\psi_0)$ for $\psi_0=12$. 
	 Result of MC simulations are shown by the solid lines
	 with the error estimates bounded by the bands of thin lines;
	 results of the linear least squares fit are shown by
	 the dashed lines and the cut-off times $t_1$ and 
	 $t_2$ are shown by circles.}
\label{F:poly_fit}
\end{figure}

From the MC results, we compute $\langle\psi(t,\psi_0)\rangle$ and
$\sigma^2(t,\psi_0)$ and obtain the time derivatives of these 
quantities by fitting a straight line to them. 
An example of obtained $\langle\psi(t,\psi_0)\rangle$ and 
$\sigma^2(t,\psi_0)$ together with the fitted lines 
is shown in Fig.~\ref{F:poly_fit}.
The fitting is performed for $t\in[t_1,t_2]$,
where $t_1$ and $t_2$ are cut-off times, 
whose choice is motivated by the following considerations.
The evolution for $t < t_1$ is neglected, since it corresponds
to ``healing" the details of our particular initialization as
we approach the one-dimensional manifold parameterized 
by $\psi$.
The one-dimensional coarse-grained description for $\psiav(t)$
becomes a valid approximation after some initial time has elapsed, 
i.e. beyond $t_1$.
The cut-off time $t_1$ was chosen by
a visual inspection of the plots and its precise choice does not
influence the results.
The relation of the multi-dimensional dynamics 
to $t_1$ will be discussed in more detail in section~\ref{S:multi_dim}.

\begin{figure}
\includegraphics[width=5in]{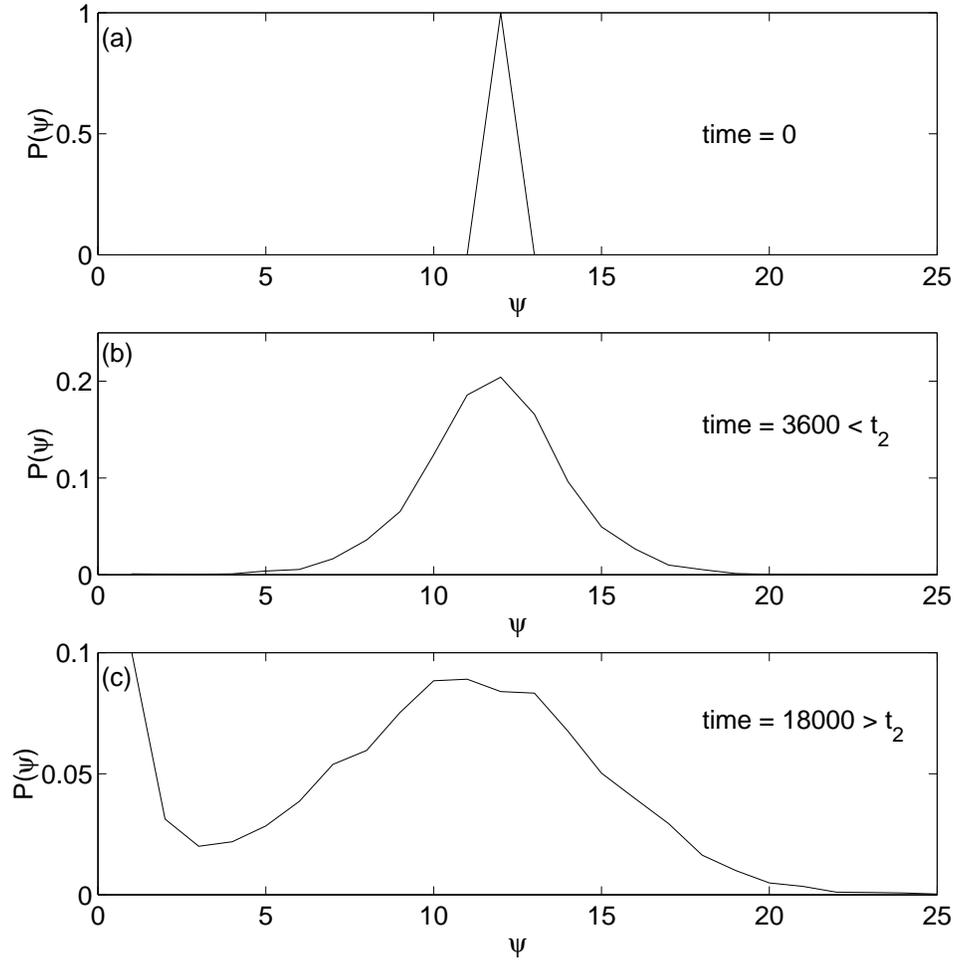}
\caption{Evolution of distribution of the cluster size $\psi$.
         Initial cluster size is $\psi_0$ = 12. 
	 (a) $\delta$-function 
         distribution at time = 0, (b) Gaussian distribution
	 at an intermediate time, and (c)  
	 bimodal distribution at a later
	 time, when a significant fraction of nuclei have disintegrated
	 into single amphiphiles, di- and tri-mers 
	 (whose dynamics is uncorrelated with $\psi_0$).}
\label{F:hist}
\end{figure}

The upper cut-off time $t_2$, corresponding to the evolution of 
the initial $\delta$-function density becoming non-Gaussian,  can 
be justified as follows.
If the initial cluster size $\psi_0$ is sufficiently small, then a
significant fraction of MC realizations will result in a complete
disintegration of the nucleus into a collection of
unconnected single amphiphiles. 
This process is illustrated in Fig.~\ref{F:hist}, which shows an
evolution of the distribution of the cluster size. 
Initially, this
is a $\delta$-function distribution. 
At some intermediate time, 
this is still well approximated by a Gaussian distribution;
we have not yet started to sample the nonlinearities of
the effective free energy away from the nominal initial point.
At some later time, when a significant 
fraction of clusters has disintegrated, the distribution starts
becoming bimodal.
This bimodality of the distribution is echoed in a nonlinear (in time)
behavior of both $\psiav(t)$ and $\sigma^2(t)$.
We hence choose the upper cut-off time $t_2$ as the time 
at which the height of the second mode is 5\% the height of
the Gaussian mode.

We observe that the introduction of the upper cut-off time $t_2$
is necessary only for relatively small clusters ($\psi_0 \le 30$). 
For larger clusters, 
$t_2$ is much larger than the simulation
time because disintegration
of a micelle into small clusters is an extremely slow process,
and the simulation does not leave the neighborhood of the bottom
of the micelle well.

\section{Results}
\label{S:results}

\begin{figure}
\includegraphics[width=5in]{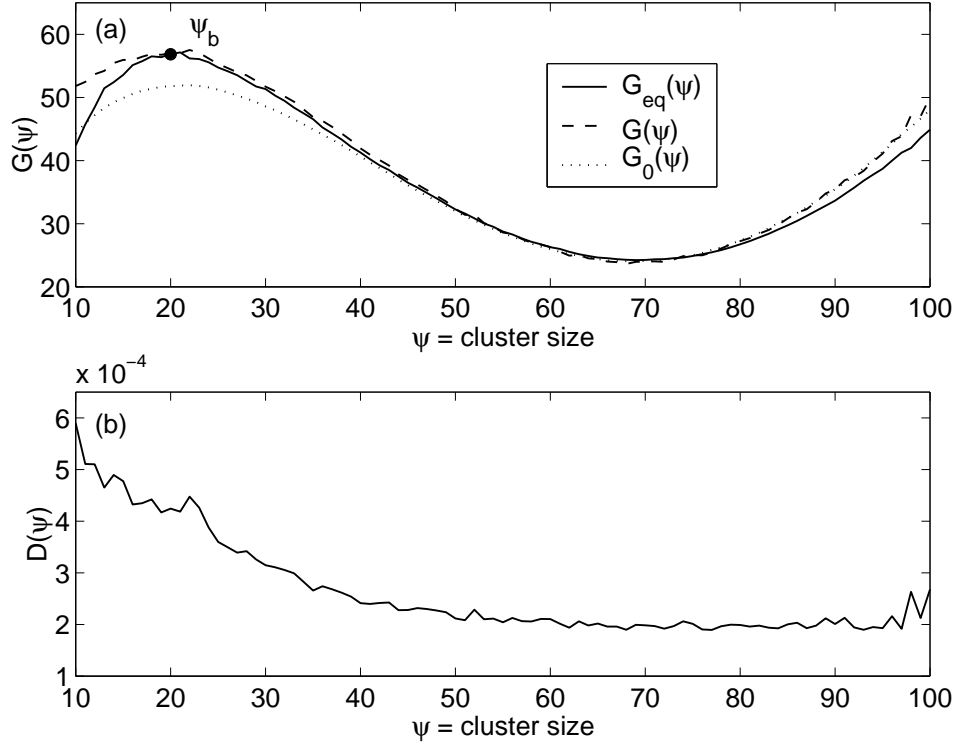}
\caption{(a) Free energy $\Geq(\psi)$ obtained from the equilibrium simulations 
             (solid line), free energy $G(\psi)$ calculated using the
	     kinetic approach and Eq. (\ref{eq:G}) (dashed line), 
	     and free energy $G_0(\psi)$ obtained from the expression
	     (\ref{eq:G0}) which neglects the spatial dependence of
	     the diffusion coefficient (dotted line);
	  (b) diffusion coefficient $D(\psi)$.}
\label{F:G}
\end{figure}

The effective free energy $G(\psi)$ obtained from the kinetic approach is compared
to the free energy $\Geq(\psi)$ 
obtained from the full-scale equilibrium simulations in Fig.~\ref{F:G}a.
We observe good agreement between the two estimates of the free energy for
the values of $\psi$ located on the right of the free energy barrier.
The discrepancy between $G(\psi)$ and $\Geq(\psi)$ 
becomes significant of the left of the
barrier. 
As will be discussed in Section~\ref{S:multi_dim}, we believe that this
discrepancy is due to the fact that the dynamics for these small $\psi$
is effectively multi-dimensional,
i.e. the timescale of the approach to the one-dimensional 
manifold is comparable to the timescale of motion along that manifold.

The effective diffusion coefficient $D(\psi)$, shown in Fig.~\ref{F:G}b,
exhibits strong position dependence near the free energy barrier.
This suggests importance of the correction to the free energy
due to multiplicative noise (see the second term in Eq. (\ref{eq:G})).
Indeed, in Fig.~\ref{F:G}a we show for comparison the free 
energy $G_0(\psi)$ obtained from
expression (\ref{eq:G0}), which neglects the multiplicative
nature of the noise.
It is clear that the discrepancy
between $G_0(\psi)$ and $\Geq(\psi)$ is significant in the
barrier region, precisely in the region of strong position
dependence of $D(\psi)$.

Drift and diffusion coefficients $v(\psi)$ and 
$D(\psi)$, obtained from the kinetic approach calculations,
can be used to obtain the {\it computational} disintegration rates of micelles.
From Eq. (\ref{eq:mean_res}), we compute the micelle disintegration
rate to be $k = 5.58 \times 10^{-9}$. 
This compares reasonably well with the 
micelle disintegration rate of $k_{1\to 0} = 7.70 \times 10^{-9}$
obtained from the full-scale MC simulations 
(see Section~\ref{S:equilibrium}).
The discrepancy is partly due to the discrepancy in free energies
on the left of the saddle point (see Fig.~\ref{F:G}a). 
In fact,  if we compute the disintegration rate using the free energy obtained 
from the equilibrium simulation and the diffusion coefficient 
obtained from the kinetic approach calculations, we obtain the
rate $k = 6.58\times 10^{-9}$, in a better agreement 
with the equilibrium result. 

Calculation of the micelle \emph{formation} rate is somewhat more
complicated because in this case one needs to examine dynamics
of very small nuclei, which cannot be described by our
one-dimensional Fokker-Planck equation parametrized by
micelle size.
In fact, a small nucleus is
indistinguishable from other small clusters in the simulation box. 
A possible solution is to match the flux 
$j_+(\psi)$ of growing cluster sizes with the flux $j_-(\psi)$ of the
decaying cluster sizes $\psi$. 
The flux $j_+(\psi)$ of nuclei emerging from the ``soup" of small clusters
can be calculated directly using short-scale simulations with initial
conditions being an empty box.
The flux $j_-(\psi)$ of disintegrating clusters can be calculated
from the Fokker-Planck equation.
In order to match these fluxes, it is required to have a 
reliable FP equation description of the cluster size evolution 
in the range of $\psi$ where the matching is expected to take place.
However, currently we observe a discrepancy between the equilibrium
simulations and the kinetically fitted single coarse variable
Fokker-Planck equation on the left
of the free energy barrier $\psi_b$, which is evidenced, e.g. by different
slopes of $\Geq(\psi)$ and $G(\psi)$ on the left of $\psi_b$ 
(see Fig.~\ref{F:G}a).
Since the matching 
should be performed for $\psi < \psi_b$, we cannot currently
estimate micelle {\it formation} rates using the 
the kinetic approach using the micelle size as a ``coarse variable".
However, the effective Fokker-Planck equation description on the
left of $\psi_b$ can be improved if 
one goes beyond the one-dimensional coarse variable model, as 
discussed in section~\ref{S:multi_dim}.

\subsection{Validity of the Fokker-Planck equation assumptions.}

\begin{figure}
\includegraphics[width=5in]{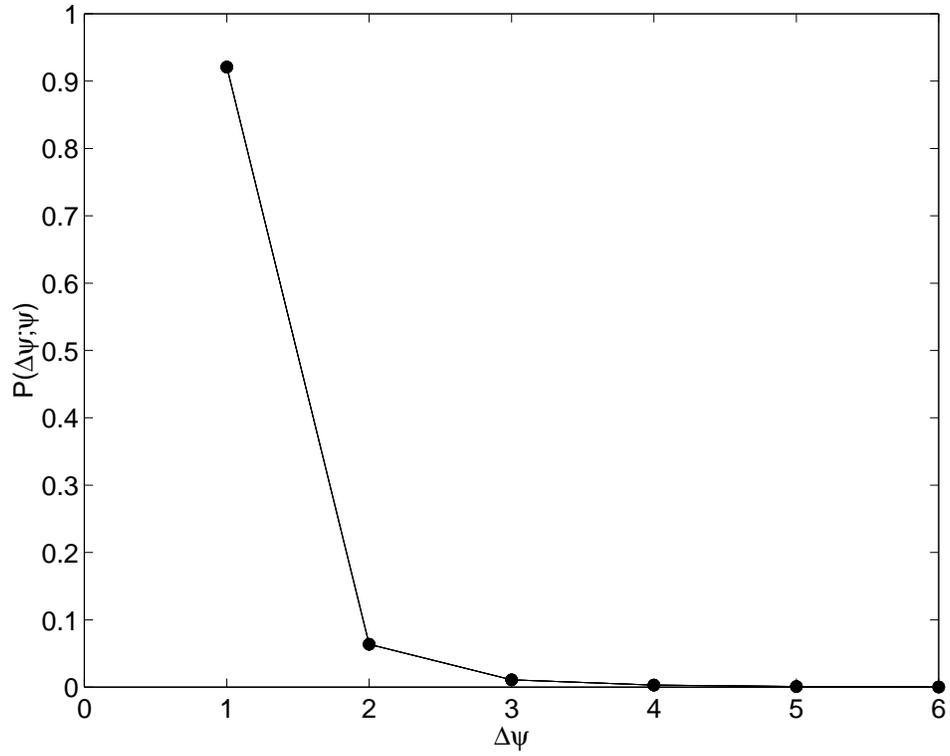}
\caption{Probability $P(\Delta\psi;\psi)$ of removal (addition) 
         of a cluster of
         size $\Delta \psi$ from (to) a nucleus of size 
	 $\psi = 10$. This probability distribution is 
	 almost identical for all nuclei sizes $\psi \ge 10$.}
\label{F:brate_prob}
\end{figure}

In this subsection, we discuss several assumptions behind the 
effective FP equation dynamics and 
demonstrate computationally that these assumptions hold
in the case of our GCMC simulations of micelle formation.
One of the assumptions implicit in the FP model is that the
cluster size $\psi$ changes gradually, i.e. removal (addition) of 
single amphiphiles (or, possibly di- and tri-mers) from (to) the
nucleus is far more probable than
spontaneous break up of a nucleus into several clusters of 
comparable size 
(spontaneous assembly of clusters into a nucleus). 
In order to check this assumption, we compute the probability 
$P(\Delta\psi;\psi)$ of removal
(addition) of a cluster of size $\Delta\psi$ from (to) the nucleus
of size $\psi$. 
We observe that, for all nuclei sizes, removal/addition of a single
amphiphile has a probability greater than 0.9 and the probabilities of 
removal/addition of larger clusters decrease monotonically with the
cluster size. 
In Fig.~\ref{F:brate_prob}, we show the
probability $P(\Delta\psi;\psi)$ for the nucleus size $\psi = 10$.
Such probability distributions are almost identical for all nucleus sizes
$\psi \ge 10$ and hence, the assumption of the gradual change of the
size of the nucleus is valid.

\begin{figure}
\includegraphics[width=5in]{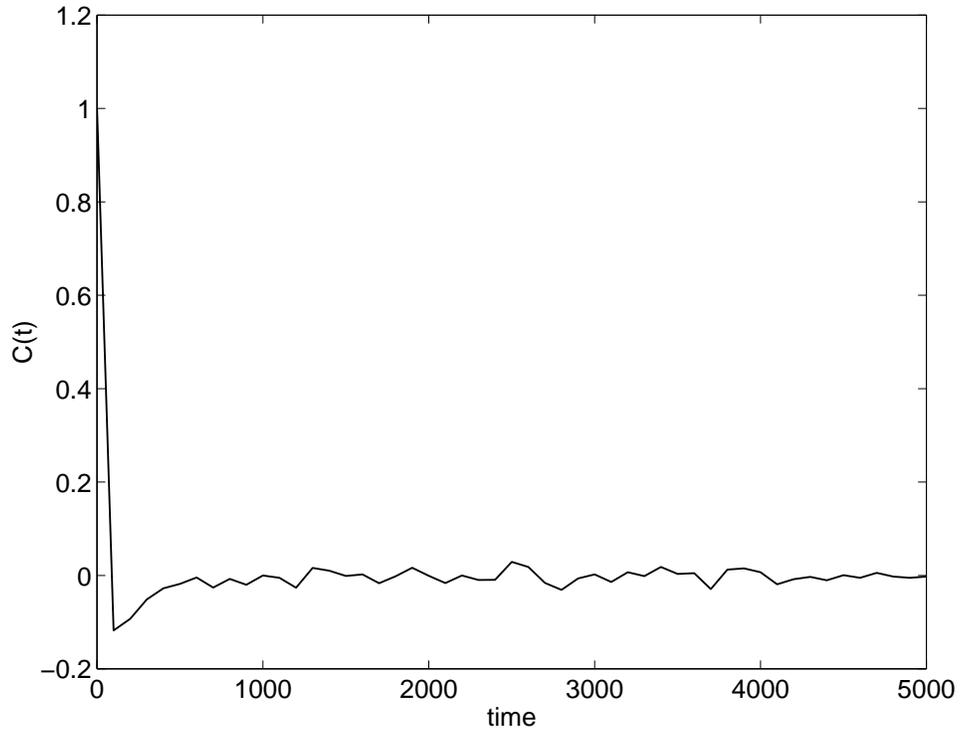}
\caption{Autocorrelation function $C(t)$ of the stochastic force $F(t)$; 
         this function is normalized so that $C(0) = 1$.
	 The shown function is computed for the initial nucleus
	 size $\psi_0=12$ and is typical for all $\psi_0 \ge 10$.}
\label{F:autocorr}
\end{figure}

Another assumption of the FP equation is that the process is Markovian.
This assumption is equivalent to the assumption (\ref{eq:FDT}) of the 
zero-correlation time of
the stochastic force in the Langevin equation (\ref{eq:LE}).
From the Langevin equation, it follows that 
the correlation time of noise coincides with that of $d\hpsi(t)/dt$, where
\begin{equation}
   \hpsi(t) = \psi(t) - \langle\psi(t)\rangle
\end{equation}
is the fluctuation of $\psi(t)$. 
Therefore, in order to estimate the correlation time of $F(t)$, we compute the
autocorrelation function $d\hpsi(t)/dt$.
The time derivative of $\hpsi(t)$ is estimated using the forward differences,
\begin{equation}
  \frac{\hpsi(t)}{dt}\Big\vert_{t = t_i} \approx 
  \frac{\hpsi(t_{i+1}) - \hpsi(t_i)}{\Delta t},
\end{equation}
where $\Delta t$ is the frequency of output in our simulations.
In the calculations of the autocorrelation functions we have used
results of the shorter MC simulation with more frequent output (see Section
\ref{S:nonequil}) and hence $\Delta t$ = 
100 MC steps.
A normalized autocorrelation function of $\hpsi(t)$ (and, hence, of $F(t)$)
for the initial nucleus size $\psi_0 = 12$ is shown in 
Fig.~\ref{F:autocorr} and is typical
for all $\psi_0 \ge 10$.
It is clear that approximating the effective stochastic practically 
by white noise is a good assumption.
Hence, the evolution of the nucleus size can be modelled by the
effective Fokker-Planck equation.

\subsection{Quality of database}
\label{S:database_quality}

Another important question that needs to be addressed is whether 
the cluster database has a sufficient number of cluster structures
in order to provide statistically accurate initial conditions
for the kinetic approach simulations.
%
This question is especially pertinent near the free energy
barrier. 
Recall that the cluster database is obtained from
the equilibrium run and the clusters are saved every 100,000
MC steps. 
%
%
Since the probability to observe a cluster near
the barrier is very low, there is a big difference in 
the number of the available cluster structures at the barrier
and in the free energy well.
The database used in most of our calculations 
was obtained from 500 realizations of 5 million MC steps and,
although there are hundreds of entries for (almost) equilibrium
micelles in the free energy well, there is as little as 3 
database entries for some cluster sizes 
near the barrier.
In order to check if this small number of initial configurations
introduces a bias into the kinetic approach simulations, 
we have added more structures to the database 
by running equilibrium MC simulations for additional 45 million 
MC steps. 
In this larger database, the smallest
number of database entries is 118.
We have repeated the calculations with this enlarged database and
obtained the same $G(\psi)$ and $\gamma(\psi)$ as we did with the
smaller database.
Hence, the kinetic approach calculations are accurate
even for small number of database entries. 
This happens because,
even if initially we place the same nucleus into several copies
of the simulation box, during the equilibration time these nuclei
will evolve into statistically different
structures.
The timescales of change of the cluster structures, as well as
biasing the equilibration by constraining the nucleus size will be
discussed in Section~\ref{S:multi_dim}. 

\subsection{Role of different dynamic rules.}
\label{S:mix}

Since MC simulations do not 
reflect the real physical dynamics and the choice of MC moves is
somewhat arbitrary, the kinetic properties 
obtained from MC simulations (such as rates of micelle formation
and disintegration) are expected to change as we change the MC
rules. 
However, if the Fokker-Planck model is valid for the 
MC ``dynamics", the equilibrium properties (such as the free 
energy), obtained from the kinetic approach should not be
affected by the change of the MC rules.

In order to confirm this, we perform MC simulations
using 9 different mixes of MC moves, which we call 
mix 0, ..., mix 8 
(mix 0 corresponds to the simulations
reported in the preceding sections).
Probabilities of different MC moves in these mixes
are listed in Table~\ref{T:mix}.
The acceptance/rejection ratios for MC moves in all 
simulations are observed to be identical.

\begin{figure}
\includegraphics[width=5in]{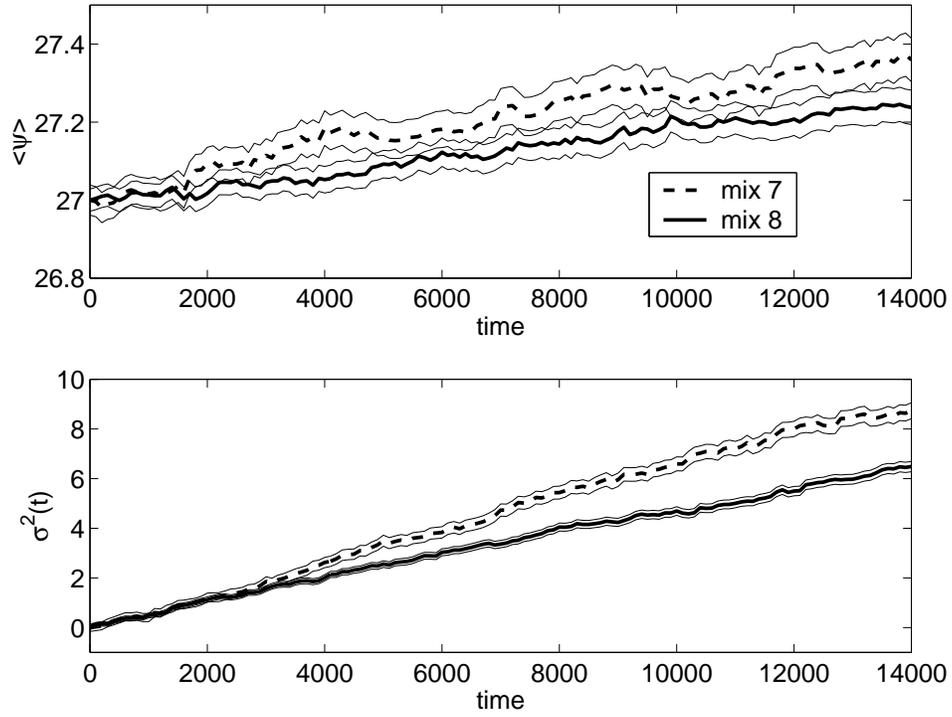}
\caption{Comparison of two short-scale simulations with different
         mixes of MC rules (mix 7 and mix 8) and the same initial 
	 nucleus size $\psi_0=27$. The error bars
	 are shown by the thin lines.
	 }
\label{F:mix78}
\end{figure}

\begin{figure}
\includegraphics[width=5in]{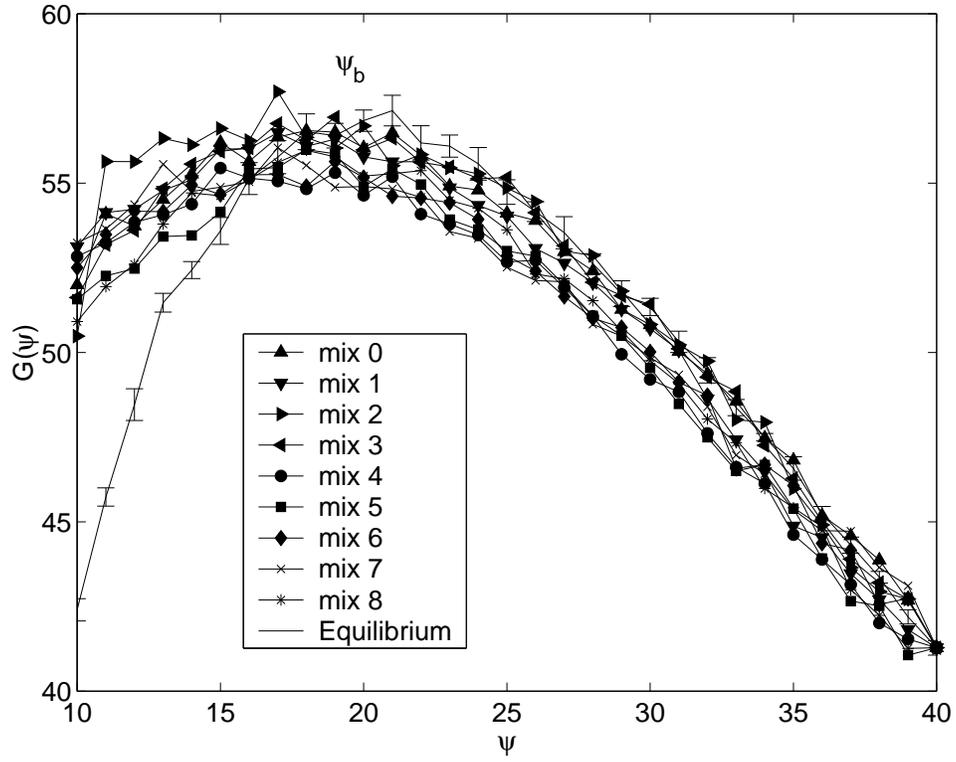}
\caption{Free energy $G(\psi)$ obtained from short-scale kinetic 
         approach MC 
         simulations with different mixes of MC moves (see
	 Table~\ref{T:mix}).}
\label{F:mix}
\end{figure}

The simulations are performed near the free energy barrier, 
with the initial nucleus size $\psi_0$ ranging from  10 to 40.
As expected, the ``dynamics" is different for 
different mixes of MC moves. 
This can be seen e.g. in 
Fig.~\ref{F:mix78} which compares evolutions of 
$\psiav(t)$ and $\sigma^2(t)$ computed from the
simulations with rules mix 7 and mix 8.
However, the free energy curves reconstructed from the short-scale
kinetic approach MC simulations agree for different MC ``dynamics", 
as seen in Fig.~\ref{F:mix}. 
%
This agreement becomes somewhat worse on the
left of the free energy barrier $\psi_b$.
The obtained results thus confirm that the Fokker-Planck model provides
an adequate description for the ``dynamics" of MC simulations of the micelle 
growth and decay.

\section{Multi-dimensional dynamics.}
\label{S:multi_dim}
 
In the previous sections, we have discussed Monte Carlo ``dynamics" of 
micelle formation assuming that the system can be accurately modeled
by a single coarse variable, namely, the aggregation number $\psi$ of a
cluster.
The aggregation number provides sufficient
information about a micelle at equilibrium. 
In fact, one can 
reconstruct the thermodynamic properties of an equilibrium micelle
of a given aggregation number using a mean-field theory 
\cite{Szleifer85a,Szleifer85b,Szleifer97}.
However, non-uniformities of \emph{non-equilibrium} clusters 
might prevent a unique specification of 
structure and physical properties of these clusters.

In this section, we explore the ``evolution" of cluster structures
and consider ``dynamics" in a two-dimensional $\psi-E$
space, where $E$ is the cluster energy. 
This variable is a useful 
probe of the cluster structure because if reflects 
how tightly the cluster is
packed: the smaller the energy, the more hydrophobic groups 
are in contact.
We note that one can choose a different variable (e.g., a radius of
gyration) to reflect the cluster structure. 
However, the specific
choice of the second coarse variable is not important: if the 
coarse-grained cluster 
dynamics are successfully parameterized by two variables, then all other variables
in our simulation become quickly slaved to the chosen two coarse variables.

\begin{figure}
\includegraphics[width=5in]{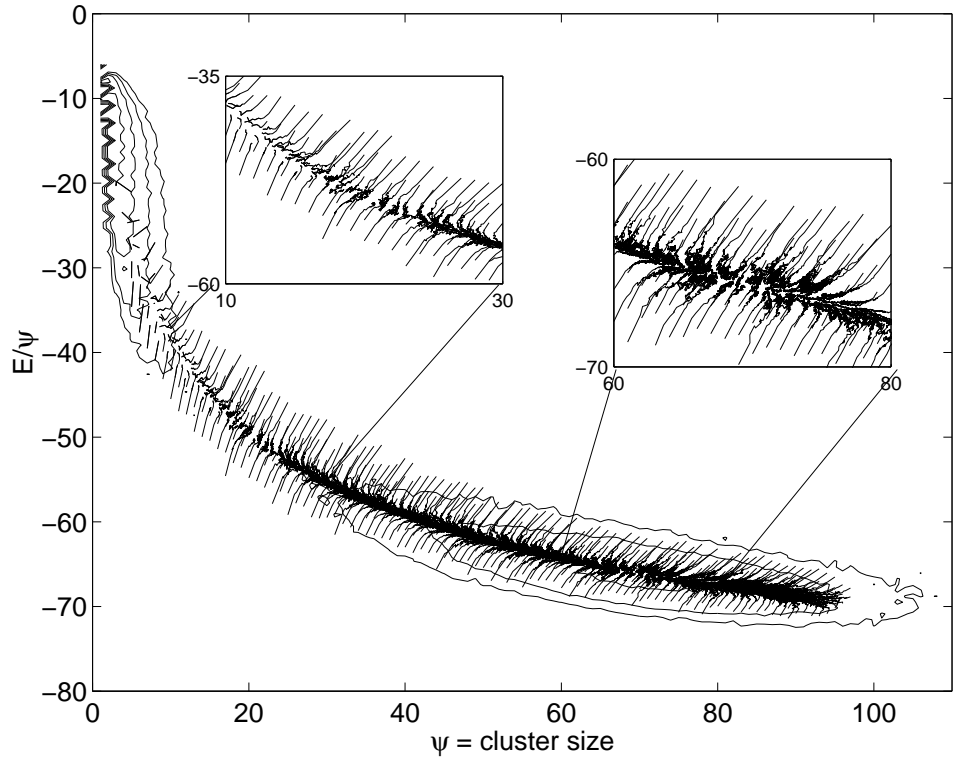}
\caption{Average trajectories in the $\psi-E/\psi$ phase space. 
         Contour plot of the equilibrium free energy $G(\psi,E/\psi)$ is also shown.
	 The insets show detailed averaged dynamics near the saddle point and 
	 the minimum of the free energy surface.}
\label{F:phase2d}
\end{figure}

It is more convenient to consider dynamics of the normalized
cluster energy $E/\psi$.
Fig.~\ref{F:phase2d} shows averaged trajectories in the 
$\psi-(E/\psi)$ phase space.
The trajectories are obtained from the
short-scale simulations described in the previous sections.
For each initial nucleus size $\psi_0$, we compute 
minimum and maximum energies
$\Emin$ and $\Emax$ of the nuclei of this size at time t=0 (i.e., just
after equilibration).
Then, the range of energies $[\Emin, \Emax]$ 
is divided into 10 equal intervals and the trajectories that 
begin at the same $\psi_0$ and in the same energy interval are 
averaged.
The free energy $G(\psi,E/\psi)$, whose contour plot is
also shown in Fig.~\ref{F:phase2d}, has been obtained from the 
full-scale equilibrium simulation, described in section
\ref{S:equilibrium}.

\begin{figure}
\includegraphics[width=6in]{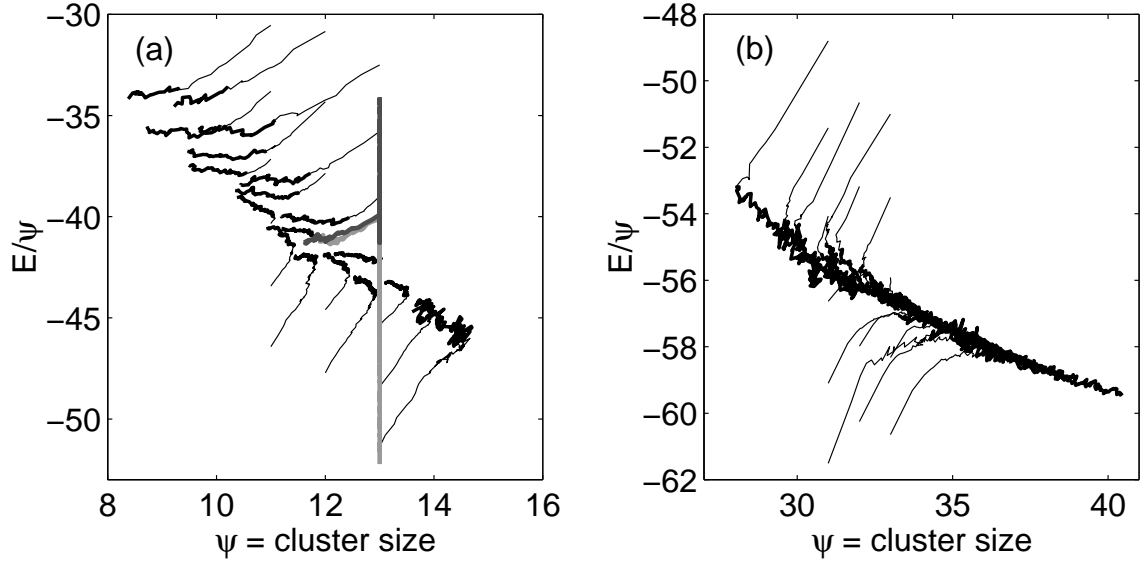}
\caption{Average phase trajectories for (a) $\psi_0 = 11, ..., 13$ (on the
         left of the free energy barrier $\psi_b = 21$) and
         (b) $\psi_0 = 31, ..., 33$ (on the right of the barrier).
	 Thin lines correspond to $t<t_1$ and thick
	 lines correspond to $t_1 < t < t_2$ (see Fig.~\ref{F:poly_fit}).
	 In plot (a), gray lines show results of simulations with the constrained at 
	 $\psi = 13$ and then released cluster size.}
\label{F:slopes2d}
\end{figure}

It is clear that the trajectories quickly approach a 
one-dimensional manifold parameterized by the cluster size 
$\psi$. 
The equilibrium micelles correspond to a stable node 
at $\psi = 69$ and the free energy barrier corresponds to a
saddle point at $\psi = 21$.
The two-dimensional dynamics provides a
clear explanation for the nonlinear behavior of $\psiav$ and
$\sigma^2(t)$ for $t < t_1$
(see Fig.~\ref{F:poly_fit} and discussion in section \ref{S:nonequil}).
For clarity, in Fig.~\ref{F:slopes2d} we plot several typical
trajectories from the complete phase portrait of 
Fig.~\ref{F:phase2d} and
indicate the part of the trajectories with $t < t_1$ by thin lines.
From these plots, it is obvious that $t_1$ corresponds to the time it takes the trajectory
to approach the one-dimensional manifold and hence, for $t < t_1$, a
one-dimensional projection $\psi(t)$ of the trajectory is a nonlinear function 
of ``time" $t$.
We emphasize that $t_1$ is not the equilibration time since the clusters 
are already equilibrated prior to computing the averages. 
However, due to statistical fluctuations in an equilibrated system, there is always a 
significant fraction of clusters away from the one-dimensional manifold. 
As Figs.~\ref{F:phase2d} and \ref{F:slopes2d} show, these clusters, 
on average, will approach this manifold.

\begin{figure}
\includegraphics[width=6in]{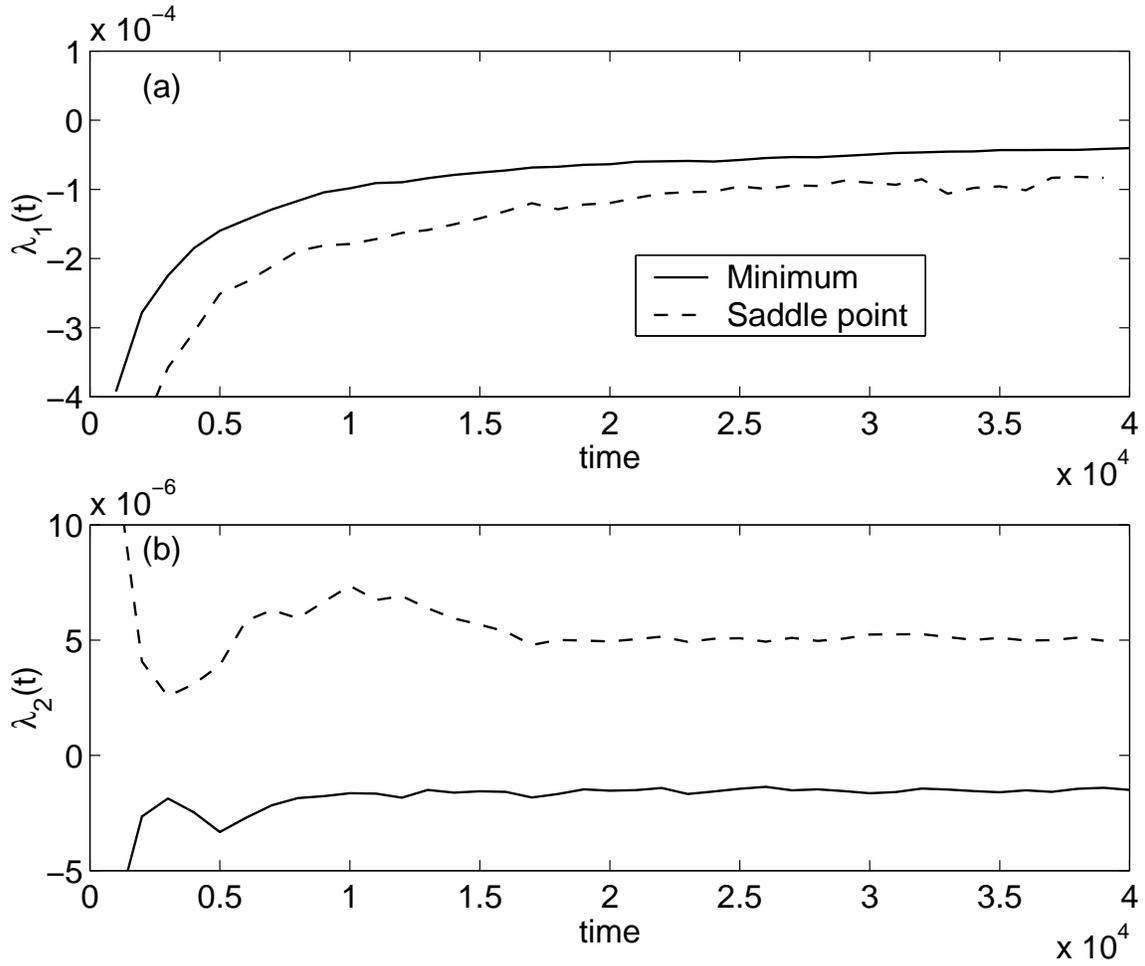}
\caption{Eigenvalues near the minimum (solid lines) and the saddle point
         (dashed lines): (a) fast eigenvalues $\lambda_1$ which 
	 characterize timescale of motion 
         towards the one-dimensional manifold and (b) slow 
	 eigenvalues $\lambda_2$
	 which characterize motion along this manifold.}
\label{F:timescale}
\end{figure}

Next, we compare the timescales of motions towards the one-dimensional manifold 
and motion along the manifold near
the critical points (saddle point and minimum) 
of the free energy. 
Near these points, the averaged dynamics can 
be approximately described by a linear homogeneous system of differential
equations,
\begin{equation}
\label{eq:lin_ODE}
   \frac{d}{dt}\left(\psiav \atop \Eav \right) = 
   A \left(\psiav \atop \Eav \right),
\end{equation}
where $A$ is a $2\times 2$ constant matrix, whose eigenvalues $\lambda_1$
and $\lambda_2$ provide
information on the timescale of motion towards and along the manifold.
In order to obtain the eigenvalues $\lambda_1$ and $\lambda_2$, we 
compute functions
\begin{equation}
   F_1(\psi_0,E_0; t) = \psiav(t)-\psi_0 \quad \mathrm{and} \quad
   F_2(\psi_0,E_0; t) = \Eav(t)-E_0,
\end{equation}
where $\psi_0$ and $E_0$ are the values of $\psiav$(t) and $\Eav$(t) at time
$t = 0$.
The eigenvalues $\mu_{1,2}(t)$ of the Jacobian $J(t)$ of the functions 
$F_1(\psi_0,E_0; t)$ and $F_2(\psi_0,E_0; t)$ are the multipliers of the
linear system (\ref{eq:lin_ODE}) and are related to the eigenvalues 
of the matrix $A$ by the following expression:
\begin{equation}
\label{eq:eig}
   \mu_j(t) = e^{\lambda_j t} - 1,\quad j = 1, 2.
\end{equation}
The Jacobian $J(t)$ is obtained from the least squares fit of
$F_1(\psi_0,E_0;t)$ and $F_2(\psi_0,E_0;t)$ to linear functions
of $\psi_0$ and $E_0$,
\begin{equation}
   F_j(\psi_0,E_0;t) \approx J_{j1}(t) \psi_0 + J_{j2}(t) E_0 + const, 
   \quad j = 1, 2.
\end{equation}
This fitting is performed using data from the averaged trajectories
$(\psiav(t),\Eav(t))$, which start from points $\psi_0$ and $E_0$ in
some neighborhood of a critical point. 
In particular, in order to 
estimate the Jacobian near the saddle point at $\psi_b$ = 21, we use
trajectories with initial nucleus size 
$\psi_0 = 18, \dots, 23$ and we use trajectories with 
$\psi_0 = 66, \dots, 74$ to estimate the Jacobian near the minimum
at $\psi_m = 69$.
The eigenvalues of matrix $A$ obtained from the multipliers $\mu_j(t)$
using Eq (\ref{eq:eig}) are plotted in Fig.~\ref{F:timescale}.
After a brief initial transient, these eigenvalues approach steady-state values.
The fast eigenvalue $\lambda_1$, shown in Fig.~\ref{F:timescale}a, 
corresponds to motion {\it towards} the manifold
and, near the saddle point, $\lambda_1 \approx -9\times 10^{-5}$ and
near the minimum, $\lambda_1 \approx -4 \times 10^{-5}$.
The slow eigenvalue $\lambda_2$, shown in Fig.~\ref{F:timescale}b, correspond to 
motion {\it along} the manifold and, near the saddle point,
 $\lambda_1 \approx 5\times 10^{-6}$
 and near the minimum, 
 $\lambda_1 \approx -1.5\times 10^{-6}$.
An order of magnitude separation of timescales appears thus to
prevail between the 
motion {\it towards} and that {\it along} the one-dimensional slow manifold.
We observe that this 
separation becomes smaller near the saddle point.

The timescale of approaching the one-dimensional manifold provides a useful measure
of how often the cluster database should be updated in order for the
saved cluster structures to be sufficiently different. 
It is reasonable to expect that within the time the 
coarse variables $\psi$ and $E$ have reached the 
manifold, the corresponding cluster structure is significantly changed.
From the eigenvalue analysis presented above, it follows that the
timescale of approaching the manifold is on the order of $10^4$ MC steps.
Hence, the frequency of the database updates used in our simulations ($10^5$ steps),
ensures that the saved structures are sufficiently different. 
Moreover, this explains why simulations with a small database 
produce results almost identical to those of simulations with a larger
database (see Section~\ref{S:database_quality}): the equilibration time of 
$2\times 10^4$ (for small clusters) and $5 \times 10^4$ (for
larger clusters) is sufficient to significantly alter the nucleus
structure and thus to provide good sampling even
if one uses a small database.
We emphasize that, although in the current work the database was obtained
from an equilibrium simulation, it can be also updated on the fly
during the kinetic simulation or possibly from an already existing
database at some nearby temperature $T$ and chemical potential $\mu$. 
In this case, estimation of the rate of change of the internal
cluster structure is crucial in order to make sure that the database
clusters become locally equilibrated.

In order to probe the multi-dimensional dynamics and approach to the 
one-dimensional manifold, we ``prepare" the micelles at the 
onset of our kinetic simulations by performing preparatory simulations 
with constrained cluster size. 
This constraint is implemented by rejecting all MC moves that
change the cluster size; approaches like umbrella sampling 
\cite{Potential} would 
also be appropriate in evolving while effectively constraining
the cluster size.

\begin{figure}
\includegraphics[width=6in]{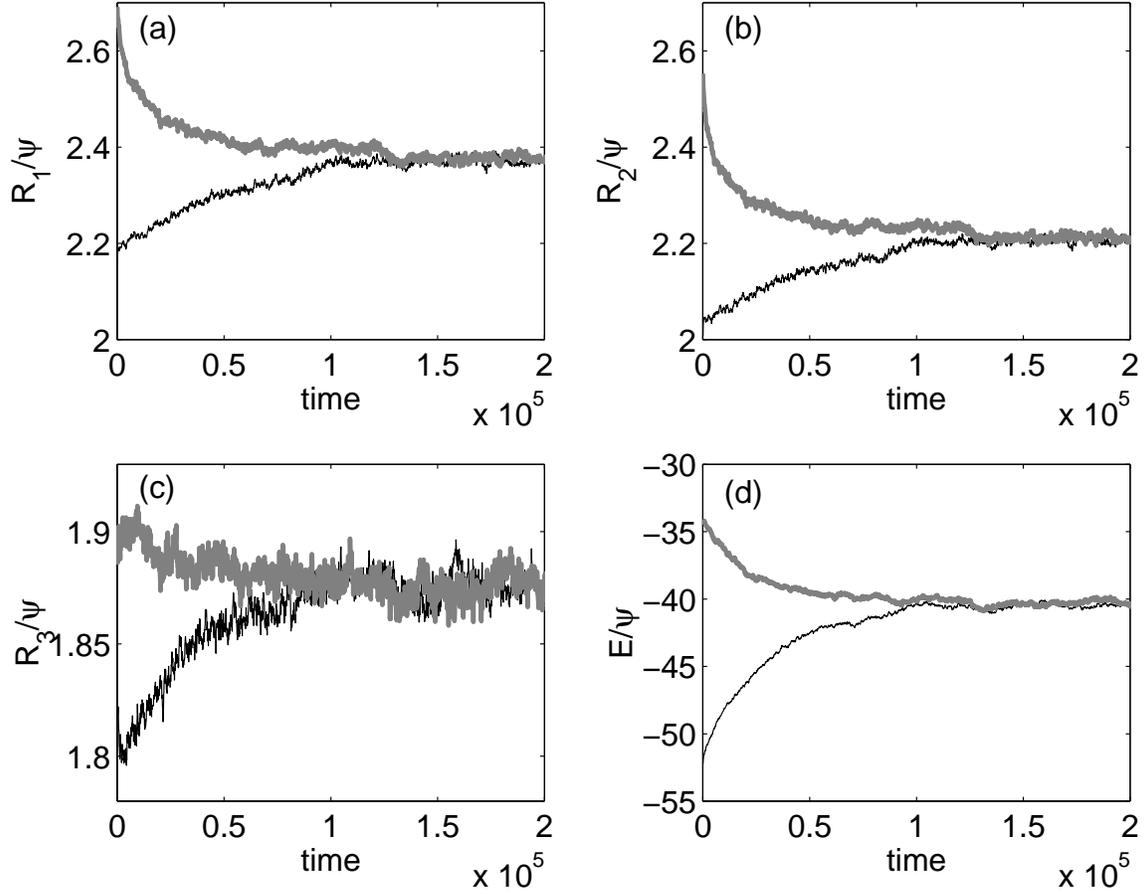}
\caption{Results of simulations with the cluster size constrained at $\psi = 13$.
         (a) through (c): 1st, 2nd, and 3rd largest normalized radii of gyration;
	 (d) normalized energy; thick gray lines show the simulation started
	 above the one-dimensional manifold and the thin black lines show the 
	 simulation started below the manifold.}
\label{F:shake_plt}
\end{figure}

We perform two simulations for the cluster size $\psi = 13$
starting above and below the one-dimensional manifold.
Evolution of cluster radii of gyration $R_1$, $R_2$, $R_3$ and energy
$E$ are averaged over 500 MC realizations and are shown in 
Fig.~\ref{F:shake_plt}.
These structural variables approach steady-state 
values that correspond to the one-dimensional manifold. 
After the cluster has approached this 
manifold, we release the constraint and let the cluster size evolve
for our kinetic simulation.
This evolution of the cluster size is shown in Fig.~\ref{F:slopes2d} by 
thick gray lines 
(the vertical lines corresponds to the constrained dynamics). 
After the cluster size is released, the phase trajectory is parallel 
to the one-dimensional manifold.

The phase trajectories shown in Fig.~\ref{F:slopes2d} also provide
an explanation for the discrepancy in the free energy $G(\psi)$
on the left of the free energy barrier $\psi_b$ (see Fig.~\ref{F:G}).
Figs.~\ref{F:slopes2d}a and \ref{F:slopes2d}b show dynamics on the left 
and on the right of the barrier, respectively.
It is clear that there is a timescale separation between the dynamics 
of approaching the manifold and motion on the manifold when the 
trajectory is on the right of the barrier. 
The timescale separation
becomes significantly smaller on the left of the barrier.
In addition, on the right of the barrier, the trajectories are much better
approximated by a singe one-dimensional manifold for $t > t_1$.
On the left of the barrier, on the other hand, the trajectories
do not quite approach a one dimensional manifold 
{\it parametrized by cluster size} and the slopes
of the trajectories which start from the same $\psi$ but 
different $E$ remain different until complete disintegration of
clusters takes place. 
Hence, in order to correctly reconstruct the free energy for these
small cluster sizes, one has to perform an analysis of the two-dimensional
dynamics.

It is interesting to notice that {\it the slope} of the one-dimensional
effective slow manifold seems to get steeper and steeper as we go towards
smaller cluster sizes. 
To deal with this, we augmented the dimension of the manifold, and added one
more coarse observable to parametrize this ``fatter" manifold.
It is conceivable that one might still be able to get a good {\it one-dimensional}
coarse description of the dynamics - but at small cluster sizes one would
need a {\it different} reaction coordinate than the cluster size;
one might still have a graph of a function above this new variable, and
not need an overall fatter two-dimensional description.
Using different order parameters at different areas of phase space, and
appropriately patching them together, is a vital area of research in 
data compression - and we are currently testing this possibility.

\section{Discussion}
\label{S:Discussion}

We have demonstrated that the Monte Carlo ``dynamics" of micelle formation
for the Larson model
can be successfully described by an effective Fokker-Planck equation,
and that the drift and
diffusion coefficients of this equation can be obtained from 
short-scale, appropriately initialized ``kinetic" simulations. 
Due to separation of timescales between the aggregation number $\psi$
of a micelle nucleus
and the coarse variables reflecting the nucleus structure (such as the
nucleus energy $E$), the coarse-grained micelle formation process can be successfully
approximated by motion on a one-dimensional manifold parameterized by
the coarse variable $\psi$. 
The separation of timescales becomes weaker for small nucleus sizes
and consideration of dynamics in a two-dimensional coarse phase
space is necessary for $\psi <\psi_b$, where $\psi_b = 21$ is the location 
of the free energy barrier.

In addition to the kinetic approach, several other ``equation-free" methods are available
that can speed up coarse-grained calculations.

\subsection{Coarse Newton method}
\begin{figure}
\includegraphics[width=5in]{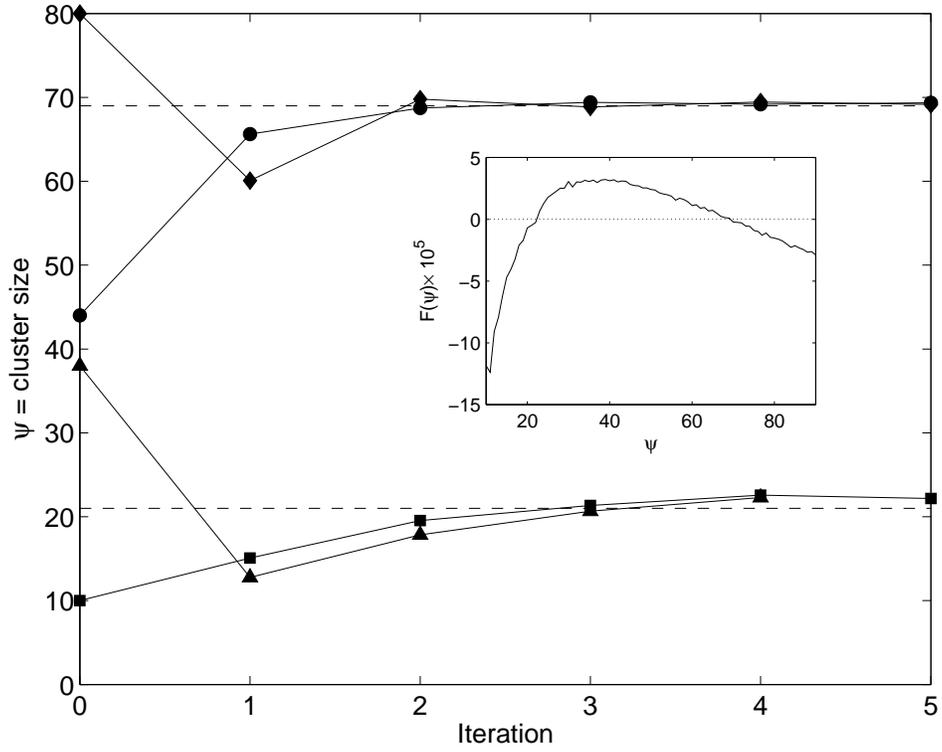}
\caption{Iterations of the Newton method starting from different 
         initial conditions. 
	 Dashed lines show the minimum at 
	 $\psi_m = 69$ and the free energy barrier at
	 $\psi_b = 21$.
	 Different symbols correspond to different simulations. 
	 Solid lines are shown to guide the eye.
	 Inset shows the function $F(\psi)$.}
\label{F:newton}
\end{figure}

The coarse Newton method as well as coarse stability and 
bifurcation analyses have been described elsewhere 
\cite{Theodoropoulos00,Theodoropoulos02,MMK02}.
In the context of the rare events problem, the
coarse Newton method can be used to obtain the location of the saddle 
point.
The Newton method was used here to locate 
zeros of the function $F(\psi_0)$ which is defined as 
the slope of $\psiav(t,\psi)$. 
In our implementation of the Newton
method, the derivative of $F(\psi_0)$ is estimated 
by fitting a straight line through points 
$F(\psi_0-1)$, $F(\psi_0)$, and $F(\psi_0+1)$.
The results of the iterations of the Newton method initialized at
different values of the coarse variable $\psi$ are shown in Fig.~\ref{F:newton}.
Depending on
the initial condition, the iterations converge either to the saddle point or
to the minimum. 
It is well known that Newton convergence requires a good initial guess.
We did, accordingly, observe that not all initial conditions
converge to a stationary point -- namely, for some points between $\psi$ = 30
and 43, the first iteration ``shoots" outside of the domain for which the
function $F(\psi)$ is defined. This is because $F(\psi)$ is very ``flat" for these
values of $\psi$ (see the inset in Fig.~\ref{F:newton}).

The function $F(\psi)$ used in our Newton method can be identified
with the driving force $f_0(\psi)$ if the Langevin equation (\ref{eq:Ito}).
The transition states (as well as the free 
energy minima) correspond to the zeros of the derivative of the
free energy $G'(\psi)$. 
In the current implementation of the coarse
Newton method, one computes the values of the coarse variable $\psi$
which correspond to the zeros of the slopes of $\psiav(t,\psi_0)$,
i.e. the zeros of the drift coefficient. 
However, as we have seen
in Section~\ref{S:theo}, the zero of the drift coefficient does not
have to coincide with the zero of the gradient of the free energy
(and it is the latter that we are after).
In fact, from equation (\ref{eq:G}) it follows that 
\begin{equation}
   G'(\psi) = 0 \quad \textrm{if and only if} \quad v(\psi) = D'(\psi).
\end{equation}
Hence, the zeros of $v(\psi)$ and $G'(\psi)$ coincide only if the
diffusion coefficient is position-independent (which is not true in the
considered case).
However, the results presented in Figs. \ref{F:G} and \ref{F:newton} 
indicate that the main correction due to the position dependence of the 
diffusivity is to the height in the free energy barrier and not the location
of the saddle point. 
Hence, we consider the results of this Newton computation representative of
the transition state; 
implementing a coarse Newton computation 
with the correction due to the state-dependent noise is straightforward.


\subsection{Coarse reverse integration}

This method has been originally developed for MD simulations in 
Ref.~\onlinecite{HK03} (see also Ref.~\onlinecite{GK03a});
after estimating the right-hand-side of an effective Langevin equation,
one can effectively reverse the time in a projective coarse Euler step 
and hence integrate {\it the coarse description} backwards in time.
In coarsely one-dimensional systems the reverse integration converges to a top
of the free energy barrier (in contrast to the forward integration which 
converges to a free energy minimum). 
In systems described by more than one macroscopic observables (reaction
coordinates), coarse reverse integration can be linked with techniques
for the construction of {\it stable manifolds} of dynamical systems
\cite{JohnsonJollyKevrekidis},
to efficiently build higher dimensional effective free energy surfaces.
%

Coarse backward integration can be readily applied to the current system.
We perform two series of reverse integration, one starting from the right
and the other starting from the left of the free energy barrier
and observe that the system indeed converges to 
the transition state.

\subsubsection{Reverse integration starting from the stable micelle}

\begin{figure}
\includegraphics[width=5in]{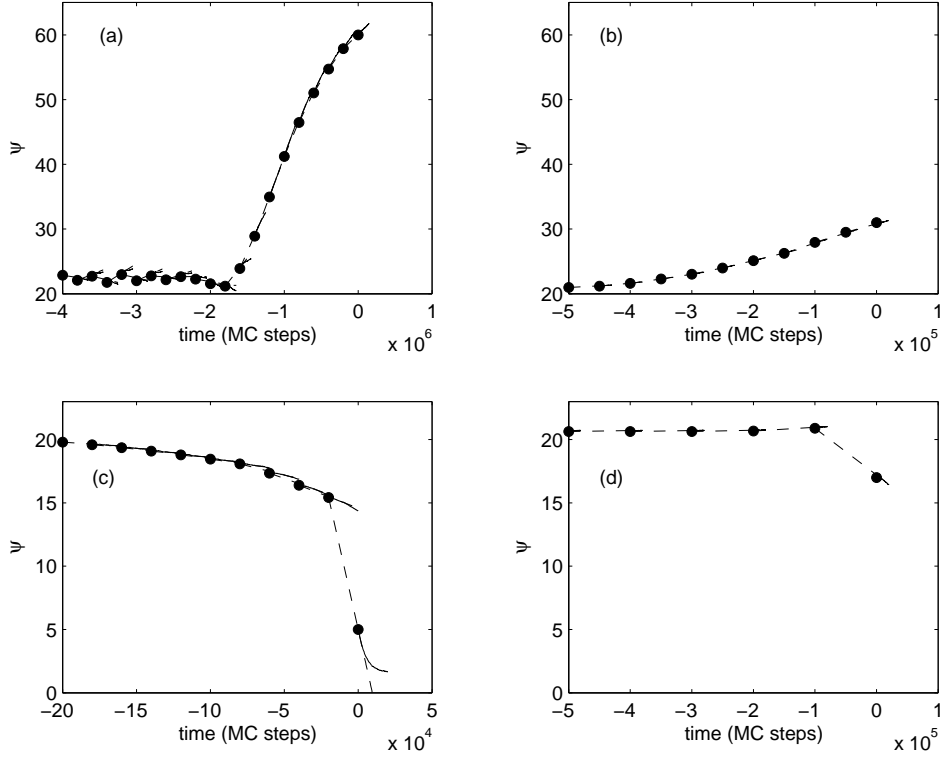}
\caption{Results of the coarse reverse integration:
(a) initial value of the coarse variable is $\psi_0=60$, 
duration of inner simulation is $\tinner = 2\times 10^5$ MC steps and 
the backward projection step is $h = -2\times 10^5$;
(b) $\psi_0=31$, $\tinner = 2\times 10^4$ MC steps, $h = -5\times 10^4$;
(c) $\psi_0=5$, $\tinner = 2\times 10^4$ MC steps, $h = -2\times 10^4$.
(d) $\psi_0=17$, $\tinner = 2\times 10^4$ MC steps, $h = -10^5$.
The solid lines show the short-scale {\it forward} simulation results and the 
dashed lines are the {\it backward} projections.
The circles show the initial conditions for the short-scale simulations.
}
\label{F:reverse_int}
\end{figure}

Results of integration starting from the right of the barrier are
shown in Fig.~\ref{F:reverse_int}a and b.
Simulations shown in Fig.~\ref{F:reverse_int}a 
start from $\psi_0 = 60$; the duration of inner simulation is 
$\tinner = 2\times 10^5$ MC steps and the {\it backward} projection step is 
$h = -2\times 10^5$.
%
The solid lines show the short-scale {\it forward} simulation results and the 
dashed lines are the reverse projections. 
The circles show the initial 
conditions for the short-scale simulations.
When the predicted (projected) state $\psi$ is at a noninteger value of the cluster size, 
given the coarse-grained nature of the computation, we use an appropriately
weighted ensemble of initial cluster sizes bracketing the desired noninteger value.
%

As $t\to -\infty$, we observe oscillations in the simulation shown in 
Fig.~\ref{F:reverse_int}a.
These oscillations are due to the large projection steps:
the integrator keeps ``overshooting" the free energy barrier.
The oscillations can be removed by the reduction in the projective 
stepsize. This is confirmed by our simulations with a smaller stepsize,
shown in Fig~\ref{F:reverse_int}b.
This simulation is performed starting from $\psi_0 = 31$; duration of the 
inner simulation is $\tinner = 2\times 10^4$ steps 
and the coarse projection step is $h = -5\times 10^4$.
The simulation converges to the location of the free energy barrier.

\subsubsection{Integration starting from the ``soup" of small clusters}

Results of the backward integration with initial conditions on 
the left of the free energy barrier are shown in Fig.~\ref{F:reverse_int}c and d.
Duration of inner {\it forward} simulation in these simulations is 
$\tinner = 2\times 10^4$.
The integration in Fig.~\ref{F:reverse_int}c is started from
$\psi_0 = 5$ and the timestep for the reverse projection is
$h = -2 \times 10^4$.
The integration approaches a steady state at $\psi = 21$, which corresponds
to the location of the free energy barrier.
As expected, the rate of convergence (measured in terms of the 
performed iterations) slows down near the barrier.
Increasing the projective stepsize to $h = -10^5$ approaches
the transition state in a smaller number of steps, as shown
in Fig.~\ref{F:reverse_int}d.
Adaptive stepsize selection (an established procedure for initial value problems,
see e.g. Ref. \onlinecite{numerical_recipes}) should in principle be used for best results.

We have therefore demonstrated that the backward timestepper with correctly 
chosen timestep converges to the location of the free energy barrier.

\subsection{Summary}

We have successfully applied the coarse kinetic approach
to the lattice Monte Carlo simulations of micelle formation. 
The approach is based on the assumption that the micelle formation
``dynamics" can be adequately described by an effective
Langevin equation model (and the corresponding Fokker-Planck 
description) for a few coarse (slow)
degrees of freedom, while treating other (fast) degrees of
freedom as a thermal noise. 
The kinetic approach,
based on short-scale
simulations with judiciously chosen initial conditions, 
then allows us to adequately reconstruct the free energy surface
and the statistical characteristics of the thermal noise.

We have shown that the micelle formation ``dynamics" can 
be parameterized by a single coarse variable, as long 
as the cluster size is sufficiently large. 
Investigations
of the system dynamics parameterized by an additional 
coarse variable (e.g. cluster energy), shows existence of
a \emph{one-dimensional} slow manifold, which is quickly
approached by the system. This separation of timescales
seizes to exist for small cluster sizes. 
This implies that
the early stages of the micelle nucleation can be characterized
by ``dynamics" on a multidimensional manifold.

We have also briefly demonstrated the application to micelle formation
of other ``equation-free" coarse numerical schemes useful in the
context of   rare event computations, such as the coarse
Newton's method and coarse reverse integration.

\begin{acknowledgements}
This work was partially supported by AFOSR and
an NSF/ITR grant. It is a pleasure to acknowledge discussions with 
Prof. C. W. Gear, Dr. G. Hummer, and Dr. M. Haataja.
\end{acknowledgements}


\newpage

\begin{table}
\begin{tabular}{l|l|l|l}
Mix   & Transfer moves & Regrowth moves & Cluster moves \\
\hline
0     &  0.5           &  0.495         & 0.005         \\
1     &  0.5           &  0.4975        & 0.0025        \\
2     &  0.5           &  0.49          & 0.01          \\
3     &  0.4           &  0.595         & 0.05          \\
4     &  0.6           &  0.395         & 0.05          \\
5     &  0.2           &  0.995         & 0.05          \\
6     &  0.8           &  0.195         & 0.05          \\
7     &  0.9           &  0.099         & 0.001         \\
8     &  0.1           &  0.88          & 0.02
\end{tabular}
\caption{Probabilities of MC moves in different mixes of
         rules used in the studies of effects
         of Monte Carlo ``dynamics" on the results of the
	 kinetic approach.}
\label{T:mix}
\end{table}

\end{document}